# Phish-IRIS: A New Approach for Vision Based Brand Prediction of Phishing Web Pages via Compact Visual Descriptors

F.C. Dalgic[1], A.S. Bozkir [2]* and M. Aydos [3]

*[1]Roketsan INC. Turkey*
*[2]Hacettepe University Department of Computer Engineering, Turkey*
*[3]Hacettepe University Department of Computer Engineering, Turkey*
*\*(selman@cs.hacettepe.edu.tr)*

*Abstract –* Phishing, a continuously growing cyber threat, aims to obtain innocent users' credentials by deceiving them via presenting fake web pages which mimic their legitimate targets. To date, various attempts have been carried out in order to detect phishing pages. In this study, we treat the problem of phishing web page identification as an image classification task and propose a machine learning augmented pure vision based approach which extracts and classifies compact visual features from web page screenshots. For this purpose, we employed several MPEG7 and MPEG7-like compact visual descriptors (SCD, CLD, CEDD, FCTH and JCD) to reveal color and edge based discriminative visual cues. Throughout the feature extraction process we have followed two different schemas working on either whole screenshots in a "holistic" manner or equal sized "patches" constructing a coarse-to-fine "pyramidal" representation. Moreover, for the task of image classification, we have built SVM and Random Forest based machine learning models. In order to assess the performance and generalization capability of the proposed approach, we have collected a mid-sized corpus covering 14 distinct brands and involving 2852 samples. According to the conducted experiments, our approach reaches up to 90.5% F1 score via SCD. As a result, compared to other studies, the suggested approach presents a lightweight schema serving competitive accuracy and superior feature extraction and inferring speed that enables it to be used as a browser plugin.

## I. INTRODUCTION

By definition, phishing is a kind of cyberattack in which the fake websites mimicking to their legitimate counterparts are exploited to trick innocent users in order to access their private and sensitive information such as username, password and bank account numbers. This kind of private information is then employed to access personal accounts for various aims that generally results illegal financial loses [1]. To date, various approaches combatting with phishing attacks have been developed. Despite of these attempts, the number of phishing attacks is continuously growing and the techniques used by phishers are evolving. According to the 2017 4th quarter phishing attacks report of Anti Phishing Working Group (APWG), more than 700.000 unique phishing attacks have been recorded during 2017 [2]. Moreover, as of October 2017, 348 target brands have been phished by scammers. Furthermore, 2017 has witnessed the exponential increase in SSL certified phishing web pages as 32% of the new phishing web pages have been equipped with SSL certificates during the 2017. This findings clearly show that, the arms race between anti phishers and scammers is continuing and the techniques of scammers are evolving to evade phishing detection mechanisms.

Rao and Pais [3] have grouped these mechanisms into 4 technical categories: (1) list based, (2) heuristics based, (3) visual similarity based and (4) machine learning based methods. Each of these methods have different pros and cons. According [4], list based attempts rely on gathering phished (black) or clean (white) URLs from various sources to provide a built-in protection mechanism often used in browsers such as Google Safe Browsing API. As [3] states, list based approaches [5] are very sensitive to URL modifications and vulnerable to *zero-day* phishing attacks. Likewise, Zhang et al. [6] addressed the limitations of whitelist based approaches by stating "as the whitelist approach is based on similarity search instead of exact matching, its detection speed is greatly affected by the feature library size and searching strategy".

As another methodology, heuristics based approaches rely on the use of features revealed from text, image or URL-specific information from legitimate and phishing websites [6]. In general, these features are then utilized by use of machine learning algorithms to form a set of heuristics for further classification of phishing websites. It is although possible to improve the classification performance with new methods Varshney et al. [6] have reported the shortcomings of heuristics based approaches in three items: (1) the need of time and computational resources for training, (2) limited ability of deployment over browsers and (3) necessity of accumulating new features once the scammers discovered the way of bypassing the system.

As is known, beyond their 1D code structure, web pages constitute 2D visual stimuli since they are rendered on web browsers. For this reason, most of the phishing attacks are based on visual deception. Therefore, recently, computer vision methods have been started to be employed to identify phishing web sites due to their inferring capabilities that are similar to biologic systems. One another important reason of preferring vision based systems is that phishers have started to deceive DOM (Document Object Model) based anti-phishing mechanisms by placing tricky contents (e.g. text like IMG tags). Thus, source contents and their corresponding 2D renderings have started to semantically vary that lowers the accuracy of recognition methods [4]. Vision based methods attempt to extract pure visual features (e.g. SIFT, SURF, HOG) that can be efficiently used to infer the brand names of the suspicious web pages. At this stage, different strategies have been demonstrated such as logo matching [7, 11], page layout similarity via HOG features [4], key point matching [1] and holistic page image matching by employing contrast context histograms [8]. Besides, Maurer and Herzner [12] have attempted to identify phishing web pages by comparing the visual signatures of the web pages that were formed by histograms of MPEG-7 and MPEG-7 like features. In contrast, Fu et al. [13] have reduced the resolution of the web pages into 100×100 and 10×10 pixels for color based feature extraction. Next, they have applied earth mover's distance (EMD) to measure the similarity of two web pages regarding to the spatially clustered and down sampled color information. Nonetheless, computational cost of EMD poses a problem for the related study. Chen et al. [8] suggested an approach which considers local visual similarities by first capturing whole screenshots of the suspicious web pages and revealing Harris-Laplace key points. Subsequently, they calculated L-CCH (lightweight Contrast Context Histogram) for each of the key points in order to be clustered based on their spatial positions via k-means algorithm [14]. Next, based on the obtained set of clusters, a similarity score can be computed that may deducing a phishing alert. However, as [14] points out, the predefined configuration of $k$ in cluster count and irregularities of key points in various web pages constitute drawbacks of the study.

Meanwhile, in [6], visual similarity based studies have been criticized due to their high computational demand compared to methods relying on text based contents. Thus, unlike pure vision based efforts, there exists various studies in literature employing different features that can be considered as "visual" since they strongly affect the visual appearance. For instance Zhang et al. [9]'s study aims to discover spatial layout similarity over spatial coordinates of page blocks which were stored in an R-tree index based database. Similarly, Jian et al. [10] employ CSS features to identify similarity between legitimate and suspicious web pages.

In this work we have proposed a novel and scalable approach to detect and recognize phishing web pages with their brand names by building machine learning models based on several easy to compute compact visual descriptors (CVD) such as SCD (Scalable Color Descriptor), CLD (Color Layout Descriptor), FCTH (Fuzzy Color and Texture Histogram), CEDD (Color and Edge Directivity Descriptor) and JCD (Joint Composite Descriptor). Further, we have suggested two visual feature extraction schemes: (1) holistic and (2) coarse-to-fine "pyramidal patches" approach. The closest work to our study in literature has been suggested by Maurer and Herzner [12] since they compute similarities of web page pairs utilizing same CVDs. However, our study presents several key differences and contributions which were listed below:

- First, we defined a data-driven and machine learning based approach employing several compact visual descriptors resulting a scalable and rapid recognition system.
- Second, we presented and examined two feature extracting schemes: (1) "holistic" – covering whole visible part of the page and (2) "pyramidal patches" – a coarse-to-fine and resolution independent scheme allowing more detailed analysis.
- Third, due to the lack of a publicly available dataset, we collected and published a mid-sized phishing dataset involving unique phishing web samples belonging to 14 targeted brands and "other" class (legitimate web pages). Our dataset totally consist of 1313 training and 1539 testing screenshots. We also made this dataset available online.
- Fourth, we conducted a comprehensive experimental study in order to evaluate our approach. The promising results show that suggested approach is efficient and competitive.

The rest of this paper is organized as follows. Section 2 briefly reviews the compact visual features employed throughout the study. Section 3 details our feature extraction methodology. Section 4 introduces our new dataset. Section 5 presents the details of the conducted experiments. Finally Section 6 discusses future directions and concludes the paper.

## II. COMPACT VISUAL DESCRIPTORS EMPLOYED

To date, due to their low computational need, compactness and ease of transportability, MPEG–7 and MPEG–7 like compact descriptors have found a widespread use in content based image retrieval and various other tasks such as object classification [15], person recognition [19] and photo annotation and retrieval [16]. Briefly, visual MPEG–7 descriptors produce 1-d representations of multimedia contents (e.g. image) by either analyzing its color, shape or texture in global manner. Yet, several MPEG-7 descriptors such as CSD (Color Space Descriptor), DCD (Dominant Colour Descriptor), SCD (Scalable Color Descriptor), CLD (Color Layout Descriptor), HTD (Homogeneous Texture Histogram) and EHD (Edge Histogram Descriptor) have been proposed and they have been successfully employed for variety of image based indexing and retrieval tasks. Furthermore, various researchers have contributed to new type of MPEG–7 like compact visual descriptors such as CEDD, FCTH and JCD [17]. To date, these descriptors have been usually preferred because of their (1) serving adequate discriminative information (2) operability even in low computing power devices and (3) generation of ultra-low sized compact signatures.

As Chieplinski [18] pointed out, color is one of most significant and easily distinguishable feature for describing visual content. Moreover, according to observations, phishing web pages involve exact or very similar colors in order to mimic to their legitimate counterparts. We, therefore, have employed 5 different color based visual descriptors (i.e. SCD, CLD, FCTH, CEDD and JCD) throughout this study. It should also be noted that, employed descriptors can take any arbitrary sized image. Due to the limited number of pages, we have provided a very short overview of the used descriptors. For further reading, readers may review [16] and [17].

The CLD divides the input image into an equal sized rectangles in a grid and computes dominant colors for each cell based on coefficients of the discrete cosine transform resulting an ultra-low length signature (i.e. 12). In other words, CLD preserves the spatial color distribution during the feature extraction. In contrast, SCD signatures are computed via encoding of Haar transform coefficients in HSV (Hue-Saturation-Value) color space enabling to have scalable representation [20]. In our implementation, SCD produces 256-d feature vectors.

Proposed by [17], FCTH and CEDD operate on same color information which is derived from two different fuzzy scheme mapping the colors of input image in a custom 24-color palette. Both of these two descriptors operate on *n* texture regions which is further separated in 24 sub areas. As stated in [17], CEDD utilizes a fuzzy version of the 5 digital filters to produce 6 texture regions whereas FCTH generates 8 texture regions via high frequency bands of the Haar Wavelet Transform. As a result, the representation of CEDD requires 54 bytes while FCTH needs 72 bytes per image. As stated above, the color information for both of these two descriptors is based on the same fuzzy scheme. Based on this fact, Chatzichristofis et al. [17] combined these two in order to create the JCD that has 7 texture areas each involving 24 sub-regions that map to color areas. According to their image retrieval and ranking based experiments, JCD produces better results.

### III. PROPOSED APPROACH

#### A. Challenges in Current State of Phishing

As explained before, the arms race between phishers and anti-phishers is continuing and the main trick of the phishers is to create deceptive web pages which mimic to their legitimate targets. Likewise, a phishing attack may get succeed only once the attacker earns trust of the innocent user. In this perspective, it can be easily deduced that human visual perception plays the key role in phishing attacks. Moreover, various tricks such as deceptive use of HTML tags makes it hard to detect via DOM based approaches. Therefore, to cope with this, computer vision (CV) based approaches have recently been studied and suggested. However, as is known, CV based approaches are generally computation intensive and an effective anti-phishing mechanism requires low latency, low memory footprint, cross-platform interoperability and high accuracy.

Currently, some of the browsers such as Google Chrome presents blacklist based phishing protectors. Nonetheless, these systems are suffering from *zero-hour* attacks which depict the instances that were not reported yet. Likewise, at present, average lifetime of a phishing web page is around 32 hours [4]. As a result, today, being robust to zero-hour attacks is a mandatory requirement.

#### B. Methodology

In the light of these facts, we investigate several compact visual descriptors in the field of phishing web page brand recognition. Thus, we first extract visual signatures from the screenshots of the web pages by two MPEG-7 (SCD and CLD) and Chatzichristofis et al.'s [17] compact descriptors (FCTH, CEDD and JCD). During this process (See Fig. 1), we not only rely on "holistic" view (i.e. single screenshot) but also do more fine-grained analysis by dividing input screenshot into equal sized 2×2, 3×3 and 4×4 grid cells which constitutes our "spatial multi-level patch pyramid" scheme. The main benefit of the spatial multi-level patch pyramid scheme is capturing the visual features regardless of the input image size and preserving spatial relationships. Hence, we no longer need to resize or crop the screenshots that causing information loss. At this point, we got inspired from the concept of *spatial pyramid matching* (SPM) which has been introduced in Lazebnik et al. work [21]. However, our approach involves a subtle difference than SPM. There exists no vector quantization and pooling stages in our proposal while SPM has local feature pooling and vocabulary generation steps as illustrated in Fig. 2.

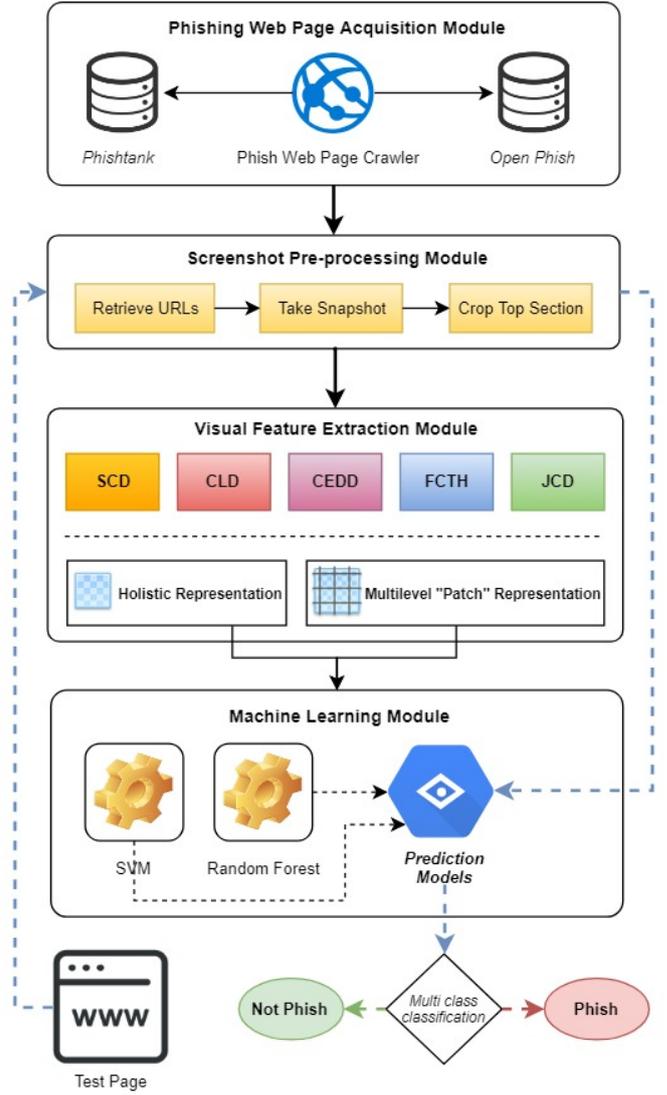

Fig. 1 Phis-IRIS System Flowchart

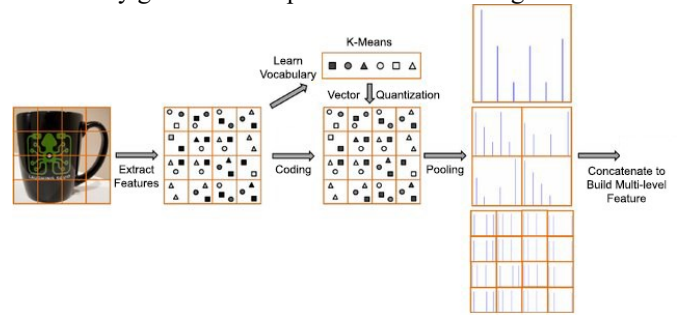

Fig. 2 Spatial pyramid based feature generation (Adopted from [22])

Following to feature vector generation of labeled data by either applying holistic or multi-level patch scheme, we have built classification models via Random Forest and Support Vector Machine methods. At this point, our work distinguishes from [12] by incorporating data driven approach. It should be noted that, our dataset includes not only heavily phished target brand classes but also "other" data. In other words, "other" class involves screenshots of legitimate web pages. Within this context, we aimed to create a machine learning model which predicts a suspicious web page's brand name or label it as legitimate.

*C. Implementation Details*

Throughout the study, we have developed two core modules named as *wrapper* and *feature extractor* respectively. We have developed these two modules in C# programming language. Our crawler is responsible for taking screenshots of phishing web pages given as URLs collected from daily Phishtank [23] and Openphish [24] reports. According to our observations, these two organizations report around 6000 phishing URLs per day. In order to process large number of web pages, we implemented our wrapper in a way that it supports multithreading which enables us taking up to 12 screenshots simultaneously. Note that, our wrapper employs Selenium Web Driver, a web automation software in order to take screenshots. During the phase of collecting screenshots, we have captured the top visible part of the web pages since this methodology has been suggested in [4, 8].

As can be seen from Fig. 3, our feature extraction tool takes an image folder and the list of selected descriptors in order to produce selected features. For the generation of all the descriptors we have employed C#.Net packages of [17]. To speed up algorithms, we implemented our feature extractor in multithreading fashion. As a result, depending on the level of multi-patching scheme and image size, our tool takes 0.21 seconds on average per image on an Intel® i7 8750H processor and 16 GB main memory equipped computer.

We employed Weka 3.8 Data Mining Software to build machine learning models. For the RF models we have not altered any parameter. Besides, we have set the cost (C) parameter of the libSVM [25] package as 40 and used RBF kernel as our basis kernel.

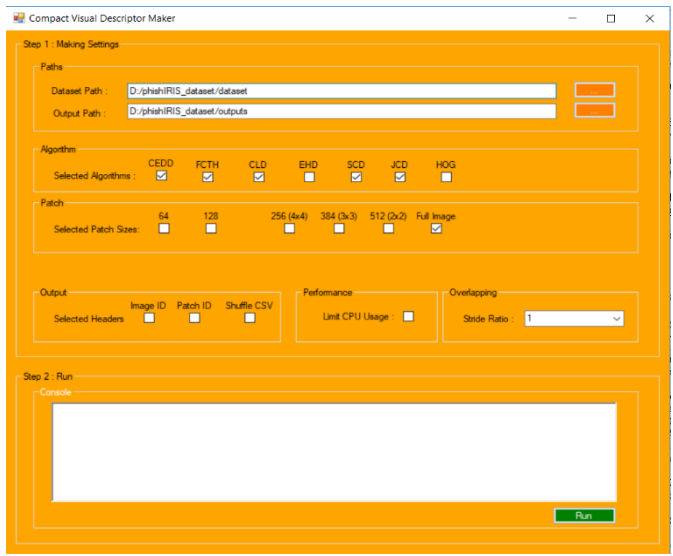

Fig. 3 Screenshot of our feature extraction module

IV. DATASET

Although there exists numerous studies in phishing literature, the number vision based approaches is limited. Moreover, according to our best knowledge, the only public phishing web page dataset involving screenshots is the [27]. However, this [27] dataset is composed of not-labeled and very limited number of samples. To this end, we have built our own dataset (i.e. "Phish-Iris-Dataset") according to our needs and made it publicly available at https://web.cs.hacettepe.edu.tr/~selman/phish-iris-dataset for further research purposes. During the dataset collection we have followed the stages described below.

First, in between March 2018 and May 2018, we have downloaded all the phishing reports of [23, 24] on daily basis in order to decide which brands should be included into our final dataset. Collected URLs were daily given to our wrapper in order to obtain their brands and screenshots. According to our observations and generated frequency reports we have listed the most phished brand names and selected the top 14 of them. Next, we have manually selected the visually unique screenshots for each brand and labeled them. As a result, we have built our corpus with non-duplicate samples. Based on our observations, we witnessed that 80% of the downloaded screenshots are just duplicates of previous samples. Therefore, finding new and novel screenshots was a challenging and time consuming task for us. Table 1 given below, presents the number of training and testing samples for each brand in our dataset.

Table 1. Brand name distributions in the corpus

| Brand Name | Training Samples | Testing Samples |
|---|---|---|
| Adobe | 43 | 27 |
| Alibaba | 50 | 26 |
| Amazon | 18 | 11 |
| Apple | 49 | 15 |
| Bank of America | 81 | 35 |
| Chase Bank | 74 | 37 |
| Dhl | 67 | 42 |
| Dropbox | 75 | 40 |
| Facebook | 87 | 57 |
| Linkedin | 24 | 14 |
| Microsoft | 65 | 53 |
| Paypal | 121 | 93 |
| Wellsfargo | 89 | 45 |
| Yahoo | 70 | 44 |
| Other | 400 | 1000 |
| Total | 1313 | 1539 |

Since most of the available web pages on Internet are legitimate, we have decided to add an "other" class which represents legitimate web pages (i.e. *negatives*). Note that, our problem at this stage becomes an open-set classification problem since the "other" class has no common color, edge structure that characterize its own class. To collect our legitimate training web pages we first listed and grabbed the screenshots of the top 300 web pages of Alexa™ by excluding the web sites of the targeted brands that we have already included. Moreover, we randomly collected another 100 and 1000 samples for training and testing respectively. As a result, we have gathered 1313 training and 1539 testing web page samples resulting totally 2852 unique screenshots covering the time frame between 4 March and 27 May of 2018.

## V. EXPERIMENTS AND RESULTS

As stated before, our main objective in this study is investigating the applicability of various compact visual descriptors in phish page detection and brand recognition task. Moreover, we also aimed to explore whether our proposed *spatial multi-level patch pyramid* feature extraction schema (i.e. multi-level-patch) contributes or not. Besides, one another goal of this study is to verify whether RF or SVM is more suitable to this problem in terms of accuracy and inference speed. According to the results, we also picked the most successful descriptor and have presented the distribution of training samples in high dimensional feature space via t-SNE method. Briefly, the method of t-SNE projects higher dimensional data into lower dimensional by preserving the individual distances between the samples in high and lower dimensional space at the same time. Note that, we have employed three types of metrics covering True Positive Rate (TPR), False Positive Rate (FPR) and F1 score (geometric mean of precision and recall). Throughout the evaluation stage, we have also calculated 5-fold cross validation results on training samples.

### A. Whole Screenshot Based Analysis

Beginning of the evaluation, we have first created the required ML models by using SVM and RF based descriptors produced via whole screenshots and measured performance of each model against testing data at the next stage. Results of the first evaluation were given in Table 2 below.

Table 2. Results of single screenshot based analysis

| Descriptor | Algorithm | # Features | 5 Fold Cross Validation | | | Testing Samples | | |
|---|---|---|---|---|---|---|---|---|
| | | | TPR | FPR | F1 Score | TPR | FPR | F1 Score |
| SCD | Random Forest | 256 | 0.842 | 0.051 | **0.844** | 0.899 | 0.101 | **0.895** |
| SCD | SVM | 256 | 0.856 | 0.038 | 0.857 | 0.885 | 0.074 | 0.886 |
| CLD | Random Forest | 12 | 0.736 | 0.062 | 0.734 | 0.814 | 0.125 | 0.814 |
| CLD | SVM | 12 | 0.699 | 0.093 | 0.697 | 0.809 | 0.190 | 0.800 |
| CEDD | Random Forest | 144 | 0.742 | 0.064 | 0.738 | 0.845 | 0.109 | 0.844 |
| CEDD | SVM | 144 | 0.695 | 0.055 | 0.705 | 0.739 | 0.078 | 0.755 |
| FCTH | Random Forest | 192 | 0.722 | 0.064 | 0.720 | 0.829 | 0.118 | 0.826 |
| FCTH | SVM | 192 | 0.669 | 0.058 | 0.670 | 0.694 | 0.084 | 0.715 |
| JCD | Random Forest | 168 | 0.767 | 0.059 | 0.765 | 0.855 | 0.114 | 0.853 |
| JCD | SVM | 168 | 0.713 | 0.051 | 0.713 | 0.728 | 0.075 | 0.748 |

Table 3. Results of Random Forest algorithm on various spatial multi-level-patch pyramid configurations

| Descriptor | # Patches | # Features | 5 Fold Cross Validation | | | Testing Samples | | |
|---|---|---|---|---|---|---|---|---|
| | | | TPR | FPR | F1 Score | TPR | FPR | F1 Score |
| SCD | 4 | 1024 | 0.835 | 0.060 | 0.838 | 0.890 | 0.138 | 0.883 |
| SCD | 9 | 2304 | 0.827 | 0.069 | 0.831 | 0.890 | 0.161 | 0.882 |
| SCD | 16 | 4096 | 0.811 | 0.076 | 0.816 | 0.879 | 0.190 | 0.869 |
| SCD | 1+4 | 1280 | 0.849 | 0.055 | **0.851** | 0.897 | 0.132 | **0.893** |
| SCD | 1+4+9 | 3584 | 0.842 | 0.059 | 0.845 | 0.895 | 0.147 | 0.889 |
| SCD | 1+4+9+16 | 7680 | 0.823 | 0.072 | 0.827 | 0.883 | 0.179 | 0.874 |
| CLD | 4 | 48 | 0.783 | 0.065 | 0.783 | 0.856 | 0.138 | 0.852 |
| CLD | 9 | 108 | 0.777 | 0.070 | 0.811 | 0.867 | 0.154 | 0.873 |
| CLD | 16 | 192 | 0.772 | 0.077 | 0.773 | 0.878 | 0.161 | 0.870 |
| CLD | 1+4 | 60 | 0.784 | 0.062 | 0.783 | 0.849 | 0.138 | 0.846 |
| CLD | 1+4+9 | 168 | 0.786 | 0.066 | 0.786 | 0.875 | 0.139 | 0.869 |
| CLD | 1+4+9+16 | 360 | 0.786 | 0.071 | 0.788 | 0.878 | 0.146 | 0.873 |
| CEDD | 4 | 576 | 0.803 | 0.056 | 0.805 | 0.875 | 0.122 | 0.871 |
| CEDD | 9 | 1296 | 0.809 | 0.063 | 0.811 | 0.890 | 0.138 | 0.884 |
| CEDD | 16 | 2304 | 0.810 | 0.067 | 0.812 | 0.875 | 0.157 | 0.868 |
| CEDD | 1+4 | 720 | 0.811 | 0.056 | 0.812 | 0.879 | 0.119 | 0.876 |
| CEDD | 1+4+9 | 2016 | 0.810 | 0.064 | 0.812 | 0.893 | 0.136 | 0.888 |
| CEDD | 1+4+9+16 | 4320 | 0.814 | 0.064 | 0.817 | 0.895 | 0.137 | 0.890 |
| FCTH | 4 | 768 | 0.804 | 0.059 | 0.804 | 0.873 | 0.118 | 0.873 |
| FCTH | 9 | 1728 | 0.795 | 0.064 | 0.795 | 0.885 | 0.135 | 0.880 |
| FCTH | 16 | 3072 | 0.801 | 0.066 | 0.802 | 0.892 | 0.139 | 0.886 |
| FCTH | 1+4 | 960 | 0.794 | 0.055 | 0.795 | 0.871 | 0.114 | 0.867 |
| FCTH | 1+4+9 | 2688 | 0.803 | 0.063 | 0.803 | 0.883 | 0.130 | 0.878 |
| FCTH | 1+4+9+16 | 5760 | 0.810 | 0.062 | 0.811 | 0.895 | 0.114 | 0.891 |
| JCD | 4 | 672 | 0.817 | 0.058 | 0.818 | 0.886 | 0.107 | 0.883 |
| JCD | 9 | 1512 | 0.804 | 0.069 | 0.806 | 0.886 | 0.145 | 0.880 |
| JCD | 16 | 2688 | 0.804 | 0.067 | 0.805 | 0.887 | 0.154 | 0.880 |
| JCD | 1+4 | 840 | 0.816 | 0.058 | 0.817 | 0.882 | 0.119 | 0.879 |
| JCD | 1+4+9 | 2352 | 0.816 | 0.061 | 0.818 | 0.887 | 0.127 | 0.882 |
| JCD | 1+4+9+16 | 5040 | 0.819 | 0.064 | 0.821 | 0.891 | 0.131 | 0.886 |

According to the Table 2, Scalable Color Descriptor modelled with RF outperforms the other models by reaching up to F1 score of 0.895. Furthermore, it can be easily seen that while RF models achieves higher F1 scores, SVM based models generate lower FPR scores on the testing dataset.

### B. Spatial Multi-Level-Patch Based Analysis

The second stage of the evaluation phase assesses and investigates the effectiveness of spatial multi-level-patch pyramid configuration. Thus, we have first designed 6 different pyramid configurations ranging from bottom only layer (e.g. 4) to 4-layers (e.g. "1+4+9+16"). More precisely, a configuration such as "1+4" refers to computing 5 different descriptors (i.e. whole screenshot + 4 equal sized rectangular regions allocating overall screenshot in total) which are then concatenated to build one long vector.

According to the results obtained with employing RF based machine learning method, we observed that the SCD slightly outperforms the other methods (See Table 3). Furthermore, considering the SCD, the best configuration has been detected as 2-level pyramid (i.e. 1+4). However, the benefit of SCD can be well understood if two other parameters (a) number of features and (b) number of patches are also considered. Likewise, the second best successful model has been detected as FCTH with 5760 features as well as 30 patches. If the computational cost of the phases like feature extraction and inferring comes into prominence, the advantage of SCD in this configuration can be better understood.

For the SVM based analysis, we have applied the same settings. The obtained results were listed in Table 4 below. The first promising finding is that SVM provides better accuracy rather than RF. Furthermore, the top predicted class (i.e. SCD with F1 score of 90.5%) not only outperforms the others but also serves less FPR.

It should also be noted that, the success of a descriptor in this study is considered in two different perspectives. In the first perspective, our criterions involve achieving higher F1 score, while not increasing the number of features much since it causes computational and storage cost. With these settings, Table 3 and Table 4 clearly states that, the SCD outperforms the others since it presents more visual discriminator power causing better distinguishing models. On the other hand, FPR score is more important for some cases. Because FPR scores denote how your model blames legitimate web pages. If we review Table 3 and Table 4, it can be seen that Joint Color Descriptor slightly wins the race. One another consistent and important finding is that, increasing number of levels in our pyramid based scheme generally resulted slightly higher F1 scores.

### C. Visualizing the Training Data

With the inclusion of "other" class into our dataset, our problem has become an open-set recognition problem. Indeed, by its nature, anti-phishing is an open set recognition problem since the recognition must not only be done against only predefined classes but it also requires the correct classification of the legitimate web pages which involve very different characteristics or patterns that were ill-sampled or not sampled in available training dataset. In order to better understand our sampling and the distribution of the 15-class dataset, we applied the t-SNE [27] method to reduce its higher dimensional to 2-d surface. To visualize the training data, we have employed related R packages and kept the perplexity value untouched. For the input, feature vectors obtained from single image via SCD has been provided.

Table 4. Results of SVM algorithm on various spatial multi-level-patch pyramid configurations

| Descriptor | # Patches | # Features | 5 Fold Cross Validation | | | Testing Samples | | |
|---|---|---|---|---|---|---|---|---|
| | | | TPR | FPR | F1 Score | TPR | FPR | F1 Score |
| SCD | 4 | 1024 | 0.853 | 0.042 | 0.854 | 0.899 | 0.084 | 0.898 |
| SCD | 9 | 2304 | 0.866 | 0.041 | 0.867 | 0.899 | 0.091 | 0.896 |
| SCD | 16 | 4096 | 0.860 | 0.043 | 0.861 | 0.899 | 0.107 | 0.895 |
| SCD | 1+4 | 1280 | 0.857 | 0.041 | 0.858 | 0.901 | 0.082 | 0.900 |
| SCD | 1+4+9 | 3584 | 0.861 | 0.042 | 0.863 | 0.906 | 0.085 | **0.905** |
| SCD | 1+4+9+16 | 7680 | 0.863 | 0.042 | **0.864** | 0.902 | 0.099 | 0.899 |
| CLD | 4 | 48 | 0.685 | 0.119 | 0.686 | 0.838 | 0.245 | 0.822 |
| CLD | 9 | 108 | 0.649 | 0.143 | 0.647 | 0.826 | 0.296 | 0.805 |
| CLD | 16 | 192 | 0.624 | 0.157 | 0.621 | 0.815 | 0.327 | 0.791 |
| CLD | 1+4 | 60 | 0.695 | 0.115 | 0.695 | 0.844 | 0.233 | 0.830 |
| CLD | 1+4+9 | 168 | 0.658 | 0.128 | 0.657 | 0.829 | 0.279 | 0.810 |
| CLD | 1+4+9+16 | 360 | 0.637 | 0.149 | 0.634 | 0.820 | 0.311 | 0.797 |
| CEDD | 4 | 576 | 0.791 | 0.042 | 0.790 | 0.786 | 0.065 | 0.798 |
| CEDD | 9 | 1296 | 0.825 | 0.039 | 0.825 | 0.821 | 0.078 | 0.827 |
| CEDD | 16 | 2304 | 0.829 | 0.037 | 0.829 | 0.832 | 0.083 | 0.836 |
| CEDD | 1+4 | 720 | 0.800 | 0.042 | 0.801 | 0.790 | 0.071 | 0.801 |
| CEDD | 1+4+9 | 2016 | 0.838 | 0.037 | 0.838 | 0.828 | 0.072 | 0.834 |
| CEDD | 1+4+9+16 | 4320 | 0.847 | 0.037 | 0.847 | 0.848 | 0.075 | 0.852 |
| FCTH | 4 | 768 | 0.756 | 0.045 | 0.758 | 0.762 | 0.067 | 0.775 |
| FCTH | 9 | 1728 | 0.794 | 0.040 | 0.794 | 0.768 | 0.081 | 0.783 |
| FCTH | 16 | 3072 | 0.814 | 0.039 | 0.814 | 0.816 | 0.072 | 0.826 |
| FCTH | 1+4 | 960 | 0.770 | 0.043 | 0.772 | 0.769 | 0.073 | 0.780 |
| FCTH | 1+4+9 | 2688 | 0.810 | 0.041 | 0.810 | 0.793 | 0.085 | 0.803 |
| FCTH | 1+4+9+16 | 5760 | 0.831 | 0.036 | 0.831 | 0.816 | 0.074 | 0.824 |
| JCD | 4 | 672 | 0.794 | 0.039 | 0.794 | 0.767 | 0.071 | 0.780 |
| JCD | 9 | 1512 | 0.823 | 0.039 | 0.824 | 0.790 | 0.084 | 0.800 |
| JCD | 16 | 2688 | 0.832 | 0.038 | 0.833 | 0.826 | 0.072 | 0.832 |
| JCD | 1+4 | 840 | 0.794 | 0.041 | 0.794 | 0.776 | 0.072 | 0.789 |
| JCD | 1+4+9 | 2352 | 0.830 | 0.039 | 0.831 | 0.814 | 0.069 | 0.822 |
| JCD | 1+4+9+16 | 5040 | 0.842 | 0.036 | 0.843 | 0.835 | 0.072 | 0.841 |

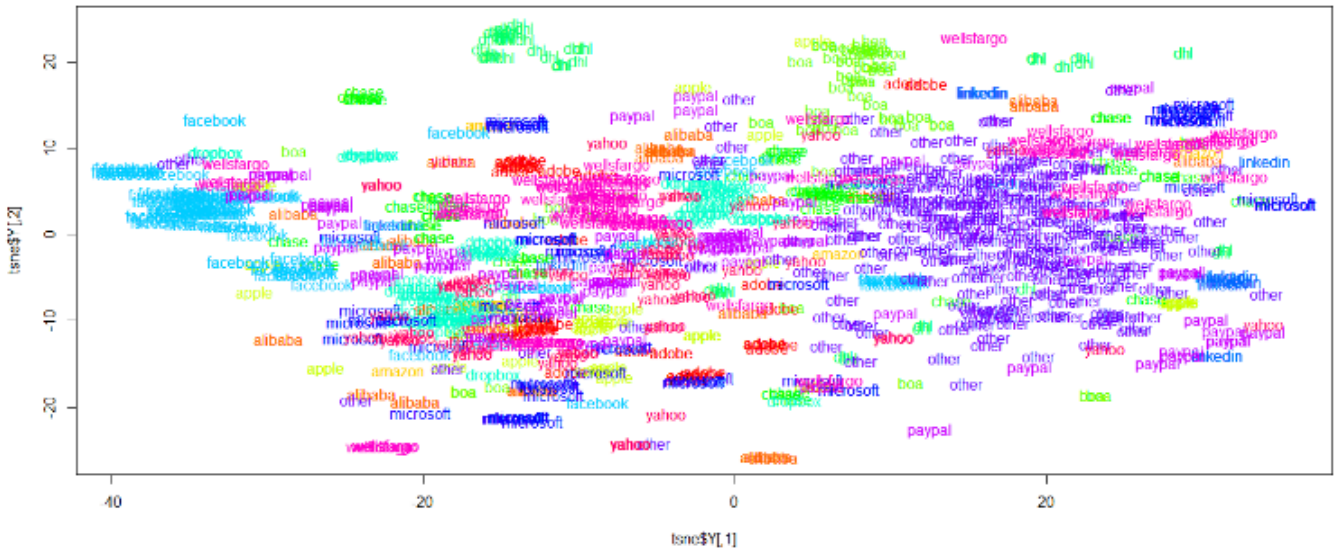

Fig. 4. t-SNE visualization of the training samples of our dataset

As can be seen from the Fig. 4, intra-class consistency is high for some classes such as 'facebook', 'boa' and 'wellsfargo'. Ideally, a perfect feature extraction scheme should separate the samples not belonging to "other" category in a way that distances between clusters must be high and number of outliers should be as low as possible. According to the Fig. 4, distributions of the classes in 2-d space point out the requirement of having better descriptors or inclusion of some other features.

## VI. COMPARATIVE STUDY

In order to compare the effectiveness of the study we compared it with [4]. In [4], phishing web pages were classified by using the features extracted with HOG (Histogram of Oriented Gradients) features. However, [4] have used 1024×1024 sized input images in order to compute valid HOG vectors. As stated before, one of the advantages of our method is being invariant to input image size. Therefore, our dataset was not initially ready to be directly processed via HOG feature extractors.

In order to get a fair comparison, we have resized the screenshots into 640×480 and applied various parameters affecting the revealed HOG features. Extracted features were than modelled with Random Forest and SVM classifiers. The comparative results are presented in Table 5. Similar to our findings, SVM based models produce much less FPR compared to RF. According to these results, SCD clearly outperforms HOG descriptors in certain conditions, while CEDD and JCD performs comparable results. Besides, it should be noted that HOG features do not involve any color specific information. They rather rely on gradients sourced from edge and corners of the input image. In this regard, the conducted comparative analysis is important since it reports whether the color or edge based information play the key role during the process of phishing web page recognition. The overall results arguably show that spatial color information has superior impact than edge or corner based information in phishing web page classification

## VII. CONCLUSION

In this paper, we aimed to build a phishing web page detection and recognition system that is scalable and robust. Hence, use of several MPEG-7 and MPEG-7 like compact color descriptors have been proposed. Moreover, along with being invariant to input image size, we also suggested the extraction and utilization of color based information via coarse- to-fine multi-level spatial patch pyramid. Due to lack of the required dataset, we collected and built a publicly available dataset for academic purposes. Conducted experiments show that our proposed approach reaches up to 90.5 % F1 score while generating 8.5% FPR and 90.6% TPR. In this regard, the models we built presents a highly effective and portable solution. The main benefit of our study is to create a lightweight solution that can be natively run on any device. As a result, compared to other studies, the suggested approach presents a lightweight schema serving competitive accuracy and superior feature extraction and inferring speed that can be used as a browser plugin or mobile device phishing protector. Nonetheless, we believe that, the proposed approach can be developed by incorporating new type of descriptors or combining the used descriptors in a late fusion fashion.

Table 5. Results of the comparative study carried out with extracted HOG Descriptors from 640×480 screenshots

| Block Size – Stride – Cell Size | Algorithm | # Features | 5 Fold Cross Validation | | | Testing Samples | | |
|---|---|---|---|---|---|---|---|---|
| | | | TPR | FPR | F1 Score | TPR | FPR | F1 Score |
| 80×80 – 40×40 – 20×20 | RF | 23760 | 0.965 | 0.059 | 0.691 | 0.822 | 0.138 | 0.820 |
| 80×80 – 40×40 – 20×20 | SVM | 23760 | 0.747 | 0.039 | 0.746 | 0.804 | 0.046 | 0.817 |
| 160×160 – 80×80 – 40×40 | RF | 5040 | 0.705 | 0.059 | 0.705 | 0.832 | 0.131 | 0.829 |
| 160×160 – 80×80 – 40×40 | SVM | 5040 | 0.744 | 0.036 | 0.742 | 0.785 | 0.049 | 0.799 |
| 320×320 – 160×160 – 80×80 | RF | 864 | 0.716 | 0.052 | 0.712 | 0.838 | 0.118 | **0.837** |
| 320×320 – 160×160 – 80×80 | SVM | 864 | 0.718 | 0.034 | 0.717 | 0.736 | 0.043 | 0.757 |

REFERENCES

[1] R. S. Rao and S. T. Ali, "A Computer Vision Technique to Detect Phishing Attack", 5th International Conference on Communication Systems and Network Technologies, pp. 596-601, 2015.
[2] APWG, Phishing Attack Trends Report 4th 2017, http://docs.apwg.org/reports/apwg_trends_report_q4_2017.pdf (Available 14.6.2018).
[3] R. S. Rao and A. R. Pais, "Detection of phishing websites using an efficient feature-based machine learning framework", *Neural Computing and Applications*, pp. 1-23, 2018.
[4] A.S. Bozkir and E. Akcapinar Sezer, "Use of HOG Descriptors in Phishing Detection", 4th International Symposium on Digital Forensic and Security, Arkansas, USA, 2016.
[5] Y. Cao, W. Han and Y. Le, "Anti-phishing based on automated individual white-list", Proceedings of the 4th ACM workshop on digital identity management, pp. 51-60, 2008.
[6] G. Varshney, M. Misra and P. K. Atrey, "A survey and classification of web phishing detection", *Security and Communication Networks*, vol. 9, pp. 6266-6284, October, 2016.
[7] K. L. Chiew, E. H. Chang, S N. Sze, W. K. Tiong, "Utilization of website logo for phishing detection", *Computers and Security*, vol. 54, pp. 16-26, 2015.
[8] K. T. Chen, C. R. Huang and C. S. Chen, "Fighting Phishing with Discriminative Keypoint Features", *IEEE Internet Computing*, vol. 13, pp. 56-63, 2009.
[9] W. Zhang, H. Lu, B. Xu and H. Yang, "Web Phishing Detection Based on Page Spatial Layout Similarity", *Informatica*, vol. 37, pp. 231-244, 2013.
[10] M. Jian, L. Pei, L. Kun, W. Tao and L. Zhenkai, "BaitAlarm: detecting phishing sites using similarity in fundamental visual features" in Proc. Intelligent Networking and Collaborative Systems 5th International Conference, 2013.
[11] G. Wang, H. Liu, S. Beccera, K. Wang, S. Belongie, H. Shacham and S. Savage, "Verilogo: Proactive Phishing Detection via Logo Recognition", UC San Diego, CS2011-0969, Tech. Rep., 2011.
[12] M. E. Maurer and D. Herzner, "Using visual website similarity for phishing detection and reporting", In CHI'12 Extended Abstracts on Human Factors in Computing Systems, 2012.
[13] A. Y. Fu, L. Wenyin and X. Deng, "Detecting phishing web pages with visual similarity assessment based on earth mover's distance (EMD)" *IEEE Trans. Depend. Secure Comput*. vol. 3, pp. 301-311, 2006.
[14] Y. Zhou, Y. Zhang, J. Xiao, Y. Wang and W. Lin, "Visual Similarity based Anti-Phishing with the Combination of Local and Global Features", 13th International Conference on Trust, Security and Privacy in Computing and Communications, 2014.
[15] N. Tsapatsoulis and Z. Theodosiu, "Object Classication Using the MPEG-7 Visual Descriptors: An Experimental Evaluation Using System of the Art Classifiers", ICANN, 2009
[16] M. Lux, "Caliph & Emir: MPEG-7 Photo Annotation and Retrieval", 17th International Conference on Multimedia, 2009.
[17] S. A. Chatzichristofis, A. Arampatzis, Y. S. Boutatlis, "Investigating the Behavior of Compact Composite Descriptors in Early Fusion, Late Fusion and Distributed Image Retrieval", *Radioengineering*, vol. 19, pp. 72-733, 2010.
[18] L. Cieplinski, "MPEG-7 Color Descriptors and Their Applications", International Conference on Computer Analysis of Images and Patterns, 2001.
[19] M. Hahnel, D. Klunder and K.F. Kraiss, "Color and Texture Features for Person Recognition", IEEE International Joint Conference on Neural Networks, 2004.
[20] J. R. Ohm, L. Cieplinski, H. J. Kim, S. Krishnamachari, B. S. Manjunath, D. S. Messing, and A. Yamada, "Color Descriptors", In: "Introduction to MPEG-7", B. S. Manjunath, P. Salembier, Th. Sikora (Eds.), pp. 187-212, John Wiley & Sons, Ltd., 2002.
[21] S. Lazebnik, C. Schmid and J. Ponce, "Beyond Bags of Features: Spatial Pyramid Matching Recognizing Natural Scene Categories", In Proc. IEEE Computer Society Conference on Computer Vision and Pattern Recognition, 2006.
[22] J. J.Thiagarajan, K. N. Karthikeyan and A. Spanias. "Local sparse coding for image classification and retrieval." *Pattern Recognition Letters*, 2012.
[23] Phishtank, [Online], Available at https://www.phishtank.com/
[24] Openphish, [Online], Available at https://openphish.com/
[25] LibSVM, [Online], Available at https://www.csie.ntu.edu.tw/~cjlin/libsvm/
[26] I. Corona, B. Biggio, M. Contini, L. Piras, R. Corda, M. Merou, G. Mureddu, D. Ariu and F. Roli, "DeltaPhish: Detecting Phishing Webpages in Compromised Websites", European Symposium on Research in Computer Security, 2017.
[27] L.J.P. van der Maaten and G.E. Hinton. "Visualizing High-Dimensional Data Using t-SNE". Journal of Machine Learning Research 9(Nov):2579-2605, 2008.